# Electron-Doping Effect on $T_c$ in the Undoped (Ce-Free) Superconductor T'-$La_{1.8}Eu_{0.2}CuO_4$ Studied by the Fluorine Substitution for Oxygen

Toshiki Sunohara, Takayuki Kawamata, Kota Shiosaka, Tomohisa Takamatsu, Takashi Noji, Masatsune Kato, and Yoji Koike

*Department of Applied Physics, Tohoku University, 6-6-5 Aoba, Aramaki, Aoba-ku, Sendai 980-8579, Japan*

We have succeeded in synthesizing electron-doped polycrystalline bulk samples of T'-$La_{1.8}Eu_{0.2}CuO_{4-y}F_y$ ($y = 0 – 0.15$) by the fluorination of undoped (Ce-free) T'-$La_{1.8}Eu_{0.2}CuO_4$ using $NH_4F$. The magnetic susceptibility measurements have revealed that the superconducting transition temperature, $T_c$, increases with increasing $y$, exhibits the maximum of 23 K at $y = 0.025$, and decreases. The dome-like dependence of $T_c$ on the doped carrier concentration in the T'-type (La,Eu)-based cuprates is explained in terms of the pairing mediated by spin fluctuations based on the *d-p* model calculation [K. Yamazaki et al., J. Phys.: Conf. Ser. **871**, 012009 (2017)].

# 1   Introduction

Since the discovery of high-$T_c$ superconductivity in the electron-doped cuprates $Ln_{2-x}Ce_xCuO_4$ (*Ln*: lanthanide elements) with the $Nd_2CuO_4$-type (so-called T'-type) structure, the superconductivity has long been believed to appear owing to both of the electron doping into the Mott-insulating mother compound T'-$Ln_2CuO_4$ and the additional reduction annealing.[1,2] The role of the reduction annealing on the appearance of superconductivity has been believed to be the removal of a small amount of excess oxygen possibly occupying the so-called apical site just above Cu in the $CuO_2$ plane.[3-5] Recently, however, it has been found that the reduction annealing adequately performed induces superconductivity even in thin films[6-8] and polycrystalline bulk samples[9,10] of the undoped (Ce-free) mother compound T'-$Ln_2CuO_4$.[11] Therefore, the electronic state and the mechanism of the superconductivity in the undoped (Ce-free) superconductor T'-$Ln_2CuO_4$ attract great interest.

As to the mechanism of the superconductivity, *d*-wave paring mediated by spin fluctuations is suggested from experimental results of the impurity effects,[12,13] μSR,[13,14] and NMR[15] in undoped (Ce-free) polycrystalline samples of T'-$La_{1.8}Eu_{0.2}CuO_4$, in which nonmagnetic $Eu^{3+}$ with a small ionic radius is partially substituted for nonmagnetic $La^{3+}$ with a large ionic one to stabilize the T'-type structure during the reduction annealing.[10] As to the electronic state of the undoped (Ce-free) superconductor, there are two possibilities. One is a strongly correlated metallic state, where the so-called charge-transfer (CT) gap between the upper Hubbard band of the $Cu3d_{x^2-y^2}$ orbitals and the $O2p$ band, which is observed in the hole-doped high-$T_c$ cuprates with the octahedral or pyramidal coordination of oxygen around Cu, is closed due to the reduction in the energies of the $Cu3d$ orbitals owing to the planar square coordination of oxygen around Cu in the T'-type structure.[16] The other is a strongly correlated metallic state with a finite CT gap and electron carriers doped into the upper Hubbard band of the $Cu3d_{x^2-y^2}$ orbitals due to oxygen defects induced by the reduction annealing.[17,18]

To clarify which is the case of the electronic state of the undoped (Ce-free) superconductor, the carrier-doping dependence of physical properties in the T'-type (La,Eu)-based cuprates is helpful. The hole-doping effect on the superconducting (SC) transition temperature, $T_c$, has been investigated by Takamatsu et al.[10,19] using Sr- and Ca-substituted samples of T'-La$_{1.8-x}$Eu$_{0.2}$M$_x$CuO$_4$ ($M$ = Sr, Ca). The magnetic susceptibility, $\chi$, measurements have revealed that $T_c$ decreases monotonically with increasing $x$, namely, with hole doping. On the other hand, the electron-doping effect was hard to be investigated, because no single-phase sample of electron-doped (Ce-substituted) T'-La$_{1.8-x}$Eu$_{0.2}$Ce$_x$CuO$_4$ could be obtained on account of the segregation of the T'-Eu$_{2-x}$Ce$_x$CuO$_4$ phase. Very recently, however, we have succeeded in synthesizing polycrystalline bulk samples of electron-doped T'-La$_{1.8}$Eu$_{0.2}$CuO$_{4-y}$F$_y$ ($y$ = 0 – 0.15) of the single phase by the fluorine substitution for oxygen. In this paper, we report how to synthesize the electron-doped samples of the single phase and the electron-doping effect on $T_c$ estimated from the $\chi$ measurements. Based on the experimental results of the $T_c$ dependence on the doped carrier concentration in the T'-type (La,Eu)-based cuprates, the electronic state and the mechanism of the superconductivity are discussed.

## 2    Experimental

Polycrystalline bulk samples of T'-La$_{1.8}$Eu$_{0.2}$CuO$_{4-y}$F$_y$ ($y$ = 0 – 0.15) were synthesized, referring to the literature on the synthesis of electron-doped T'-type cuprates by the fluorination.[20-22] First, powdered samples of T'-La$_{1.8}$Eu$_{0.2}$CuO$_4$ were prepared by the low-temperature synthesis method.[23] That is, samples of La$_{1.8}$Eu$_{0.2}$CuO$_4$ with the K$_2$NiF$_4$-type (so-called T-type) structure were prepared by the conventional solid-state reaction method, using stoichiometric amounts of La$_2$O$_3$, Eu$_2$O$_3$, and CuO powders. Then, both the powdered sample of T-La$_{1.8}$Eu$_{0.2}$CuO$_4$ and CaH$_2$ powder were sealed in an evacuated glass tube and heated at 300°C for 12 h, so that the powdered sample was reduced by H$_2$ gas produced by the decomposition of CaH$_2$ to be La$_{1.8}$Eu$_{0.2}$CuO$_{3.5}$ with the Nd$_4$Cu$_2$O$_7$-type structure. Then, powdered samples of T'-La$_{1.8}$Eu$_{0.2}$CuO$_4$ were prepared by heating La$_{1.8}$Eu$_{0.2}$CuO$_{3.5}$ at 600°C for 12 h in flowing gas of O$_2$. Secondly, the obtained powdered sample of T'-La$_{1.8}$Eu$_{0.2}$CuO$_4$ was mixed with NH$_4$F powder in the molar ratio of T'-La$_{1.8}$Eu$_{0.2}$CuO$_4$ : NH$_4$F = 1 : $y$ ($y$ = 0 – 0.15) and then pressed into pellets. The pellets were sealed in an evacuated glass tube and heated at 350°C for 10 h, so that polycrystalline samples of T'-La$_{1.8}$Eu$_{0.2}$CuO$_{4-y}$F$_y$ (called as-prepared samples) were obtained. To remove excess oxygen and obtain SC samples of T'-La$_{1.8}$Eu$_{0.2}$CuO$_{4-y}$F$_y$ (called reduced samples), the reduction annealing of the as-prepared samples was performed at 650°C for 24 h in vacuum at a pressure of ~3×10$^{-5}$ Pa. As-prepared and reduced samples of T'-La$_{1.8}$Eu$_{0.2}$CuO$_{4-y}$F$_y$ were characterized by the powder x-ray diffraction using Cu K$_\alpha$ radiation. The fluorine content in the samples was analyzed by the electron-probe x-ray microanalysis (EPMA). To detect the SC transition, $\chi$ measurements were carried out using a SC quantum interference device (SQUID) magnetometer (Quantum Design MPMS). The $\chi$ was measured to investigate the magnetism in the normal state also.

## 3 Results and discussion

*3.1 Characterization of as-prepared and reduced samples of T'-La$_{1.8}$Eu$_{0.2}$CuO$_{4-y}$F$_y$*

Figures 1(a) and 1(b) show the powder x-ray diffraction patterns of as-prepared and reduced samples of T'-La$_{1.8}$Eu$_{0.2}$CuO$_{4-y}$F$_y$ ($y$ = 0 – 0.15), respectively. It is found that all the samples are of the single phase with the T'-type structure and that the peak width is almost independent of $y$ in both as-prepared and reduced samples of T'-La$_{1.8}$Eu$_{0.2}$CuO$_{4-y}$F$_y$, indicating that the crystal quality does not change by the fluorine substitution for oxygen. Moreover, it is found that the peak width in the reduced samples is sharper than that in the as-prepared ones. This may be due to the reduction annealing at a temperature as high as 650°C enhancing the crystal quality.

The lattice constants $a$ and $c$ estimated from these diffraction patterns are shown in Figs. 2(a) and 2(b), respectively. It is found that the $a$-axis ($c$-axis) length increases (decreases) with increasing $y$ for both as-prepared and reduced samples, which is similar to the change of lattice constants by the fluorination in T'-Nd$_2$CuO$_4$.[20-22] The decrease in the $c$-axis length with increasing $y$ is interpreted as being due to the substitution of F$^-$ ions for O$^{2-}$ ones, because the ionic radius of F$^-$ is smaller than that of O$^{2-}$. On the other hand, the increase in the $a$-axis length with increasing $y$ is interpreted as being due to the electron doping into the CuO$_2$ plane. If fluorine were substituted for oxygen in the CuO$_2$ plane, the $a$-axis length should decrease with increasing $y$, which is contrary to the result. Therefore, fluorine is concluded to be substituted for oxygen around (La,Eu) in the so-called blocking layer.[20] Moreover, it is found that the $a$-axis ($c$-axis) lengths in the reduced samples are larger (smaller) than those in the as-prepared ones.

This is reasonably interpreted as a result of the removal of excess oxygen at the apical site by the reduction annealing.  That is, the increase in the $a$-axis length in the reduced samples is due to the increase in the Coulomb repulsion between $(La,Eu)^{3+}$ ions in the blocking layer and/or the electron doping into the $CuO_2$ plane caused by the removal of excess oxygen. The decrease in the $c$-axis length in the reduced samples is simply attributed to the removal of excess oxygen with a finite size.  Accordingly, these results of the lattice constants indicate the success of the substitution of fluorine for oxygen in the blocking layer and support the removal of excess oxygen by the reduction annealing.

The inclusion of fluorine in the reduced samples was confirmed by EPMA also. As shown in Fig. 3, it turns out that the nominal contents of fluorine are included in the samples within the experimental accuracy.

As mentioned in Sec. 1, nonmagnetic $Eu^{3+}$ is partially substituted for nonmagnetic $La^{3+}$ in the present samples to stabilize the T'-type structure.  Since it is generally possible that $Eu^{3+}$ changes to magnetic $Eu^{2+}$, we measured the temperature dependence of $\chi$ for the present samples.  Consequently, no Curie-like behavior due to magnetic $Eu^{2+}$ was observed as expected.  Therefore, Eu in the present samples is nonmagnetic, and the present samples of T'-$La_{1.8}Eu_{0.2}CuO_{4-y}F_y$ are very suitable for the study of the magnetism of Cu in the T'-type cuprates.

*3.2 Superconductivity and phase diagram*

To investigate the SC properties, the temperature dependence of $\chi$ was measured in a magnetic field of 10 Oe on zero-field cooling (ZFC) and field cooling (FC).  It has been found that as-prepared samples of T'-$La_{1.8}Eu_{0.2}CuO_{4-y}F_y$ are not SC above 2 K, which is common in the T'-type cuprates. As for reduced samples of T'-$La_{1.8}Eu_{0.2}CuO_{4-y}F_y$, on the other hand, SC transitions are observed in $y = 0 - 0.125$, as shown in Fig. 4.  The reduced sample of $y = 0.15$ is not SC above 2 K.  The SC volume fraction is estimated as 6% - 19% for these SC samples by the simple comparison of the $\chi$ value at the lowest temperatures on ZFC with the magnitude of perfect diamagnetism.  Taking into account the fact that these samples are finely powdered, it is said that the superconductivity is of the bulk and moreover appears in the almost whole region of these samples.  Values of $T_c$, defined at the onset temperature of the diamagnetic signal, are plotted in Fig. 5 together with those of hole-doped T'-La$_{1.8-x}$Eu$_{0.2}$M$_x$CuO$_4$ (M = Sr, Ca) reported by Takamatsu et al.[10,19]  It is found that $T_c$ increases with increasing $y$, exhibits the maximum of 23 K at $y$ = 0.025, and decreases.  Taking into account $T_c$'s of hole-doped T'-La$_{1.8-x}$Eu$_{0.2}$M$_x$CuO$_4$ (M = Sr, Ca), the carrier-doping dependence of $T_c$ is found to be dome-like.

Here, we discuss the dome-like dependence of $T_c$ on the carrier doping, based on the two possibilities in the electronic state of the undoped (Ce-free) superconductor mentioned in Sec. 1.  In the case of a strongly correlated metallic state without CT gap, the dome-like dependence of $T_c$ is explained as being due to the degree of the nesting of the Fermi surface.  That is, the best nesting induces the largest spin fluctuations at $y$ = 0 – 0.05, where the electronic band forming the Fermi surface is almost half-filled, leading to $d$-wave superconductivity with the highest $T_c$.  As shown by the dashed line in Fig. 5, in fact, a similar dependence of $T_c$ on the doped electron concentration has been obtained by Yamazaki et al.[24] from the calculation based on the spin-fluctuation-mediated pairing in the effective two-band model derived from the $d$-$p$ model, where the energy difference between the upper Hubbard level of the Cu$3d_{x^2-y^2}$ orbitals and the O$2p$ level, $\Delta$, is smaller than 1.5 eV.  In this case, the reduction annealing for the appearance of superconductivity is regarded as playing a role of the removal of excess oxygen operating to localize carriers in the CuO$_2$ plane.  In the case that the electronic state of the undoped (Ce-free) superconductor is a strongly correlated metallic one with a finite CT gap and electron carriers doped into the upper Hubbard band of the Cu$3d_{x^2-y^2}$ orbitals, on the other hand, the dome-like dependence of $T_c$ is explained as similarly as in the case of the hole-doped high-$T_c$ cuprates. That is, the Mott-insulating state is regarded as existing in T'-La$_{1.8-x}$Eu$_{0.2}$M$_x$CuO$_4$ (M = Sr, Ca) with $x$ > 0.15.  The reduction annealing is regarded as operating to make oxygen defects in the CuO$_2$ plane so as to dope electron carriers in the upper Hubbard band.[25,26]  To clarify which is the case of the electronic state of the undoped (Ce-free) superconductor, accordingly, the precise analysis of the oxygen content in both as-prepared and reduced samples is crucial, though it has been rather hard so far.  Moreover, the carrier-doping dependence of physical properties other than $T_c$ in the T'-type (La,Eu)-based cuprates is helpful also.

## 4. Summary

We have successfully synthesized polycrystalline bulk samples of electron-doped T'-La$_{1.8}$Eu$_{0.2}$CuO$_{4-y}$F$_y$ ($y$ = 0 – 0.15) by the fluorination of the undoped (Ce-free) superconductor T'-La$_{1.8}$Eu$_{0.2}$CuO$_4$ using NH$_4$F. The lattice constants estimated from the powder x-ray diffraction patterns have suggested that fluorine is substituted for oxygen in the blocking layer and that the reduction annealing operates to remove excess oxygen. The $\chi$ measurements have revealed that $T_c$ increases with increasing $y$, exhibits the maximum value of 23 K at $y$ = 0.025, and decreases. Taking into account $T_c$'s of hole-doped T'-La$_{1.8-x}$Eu$_{0.2}$M$_x$CuO$_4$ ($M$ = Sr, Ca), the carrier-doping dependence of $T_c$ has been found to be dome-like. Assuming that the electronic state of the undoped (Ce-free) superconductor is a strongly correlated metallic one without CT gap, the dome-like dependence of $T_c$ is explained as being due to the best nesting of the Fermi surface at $y$ = 0 – 0.05. This is supported by the calculation by Yamazaki et al.[24] based on the spin-fluctuation-mediated pairing in the effective two-band model derived from the $d$-$p$ model. To decide the electronic state, however, the precise analysis of the oxygen content and/or the investigation of the carrier-doping dependence of physical properties other than $T_c$ in the T'-type (La,Eu)-based cuprates is necessary.

**Acknowledgments**

We are grateful to H. Tsuchiura, K. Yamazaki, M. Ogata, and T. Adachi for the useful discussion. We are indebted to A. Arakawa for her aid in the EPMA measurements. This work was supported by JSPS KAKENHI Grant Number 17H02915.

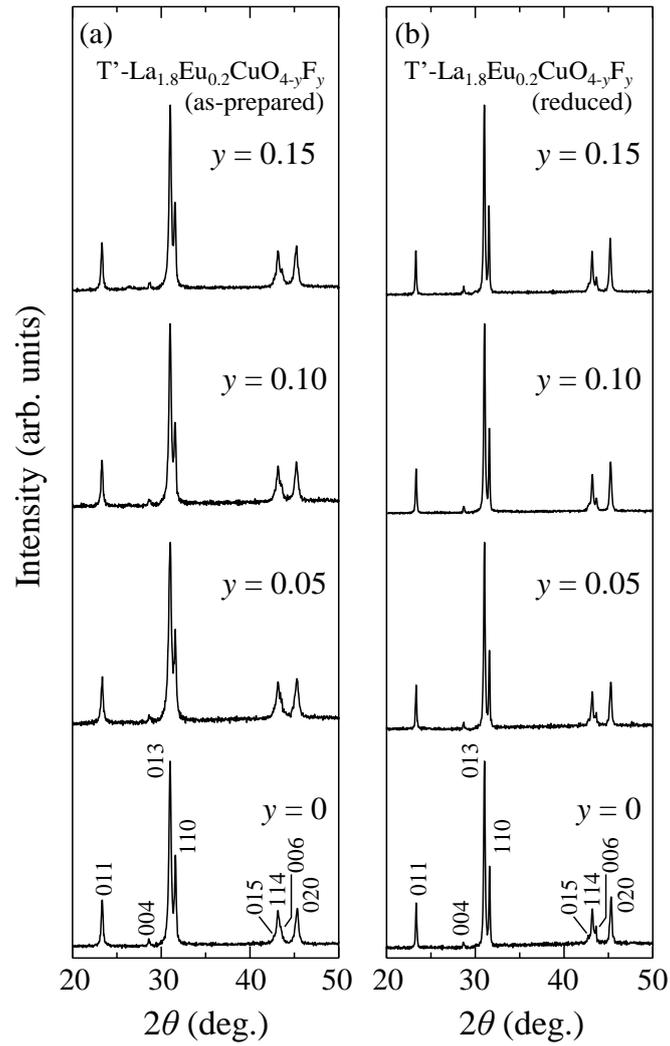

Fig. 1. Powder x-ray diffraction patterns of (a) as-prepared and (b) reduced samples of T'-$La_{1.8}Eu_{0.2}CuO_{4-y}F_y$ ($y = 0 - 0.15$).

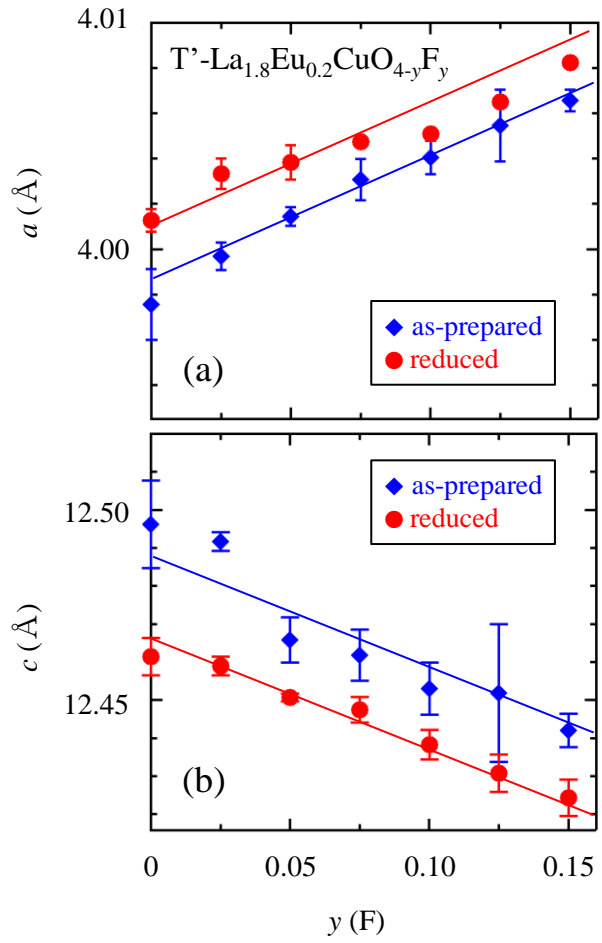

Fig. 2. (color online) Fluorine content $y$ dependence of the lattice constants (a) $a$ and (b) $c$ for as-prepared (diamonds) and reduced (circles) samples of T'-La$_{1.8}$Eu$_{0.2}$CuO$_{4-y}$F$_y$. Solid lines are guides to the eyes.

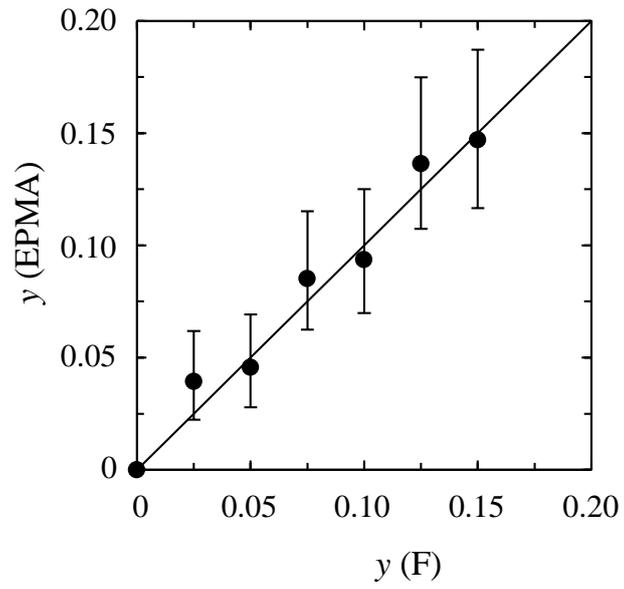

Fig. 3. Dependence of the fluorine content $y$(EPMA) estimated by EPMA on the nominal fluorine content $y$ in reduced samples of T'-$La_{1.8}Eu_{0.2}CuO_{4-y}F_y$.

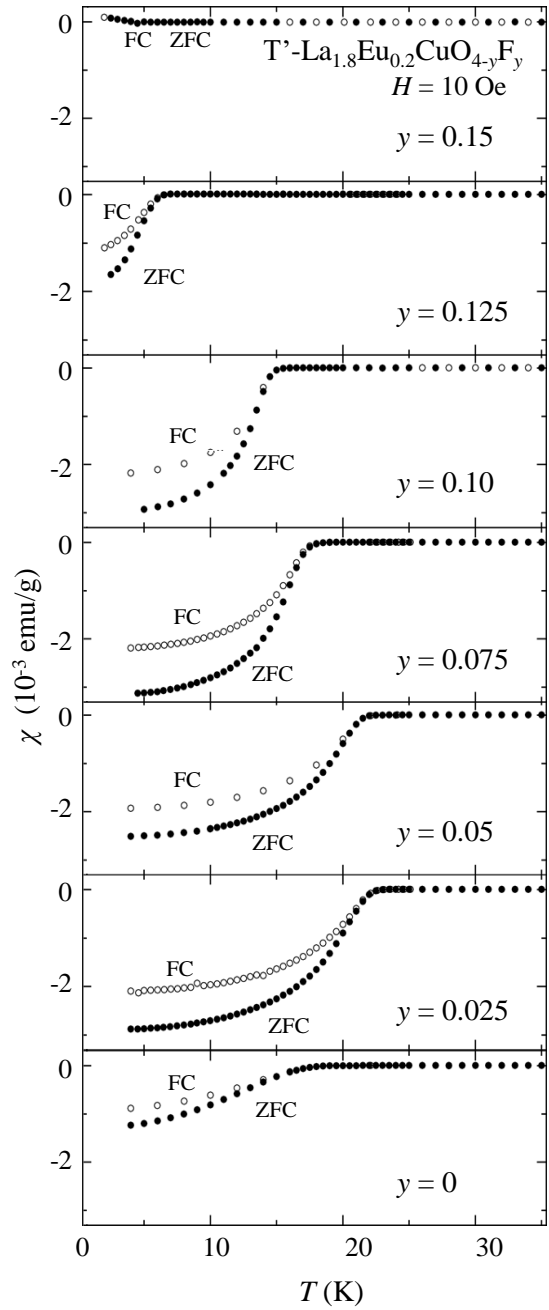

Fig. 4. Temperature dependences of the magnetic susceptibility, $\chi$, for reduced samples of T'-$La_{1.8}Eu_{0.2}CuO_{4-y}F_y$ measured in a magnetic field of 10 Oe on zero-field cooling (ZFC) (closed circles) and field cooling (FC) (open circles).

13 / 14

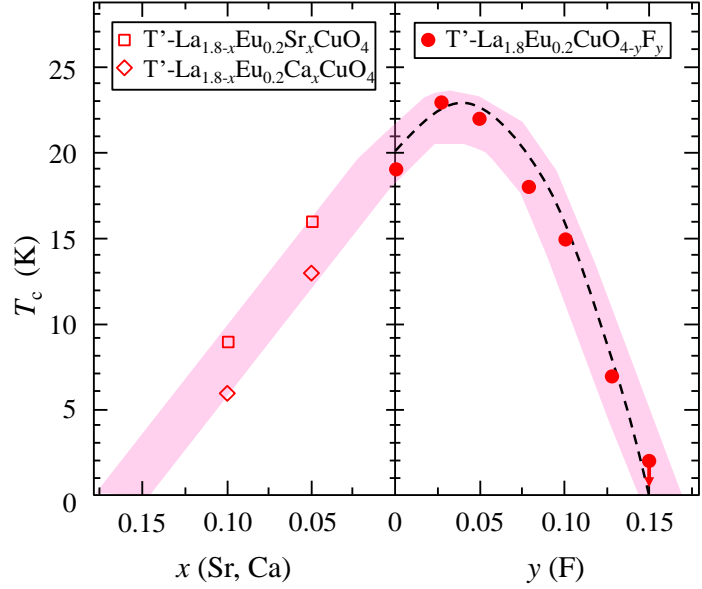

Fig. 5. (color online) Fluorine content $y$ dependence of $T_c$ in the electron-doped superconductors T'-$La_{1.8}Eu_{0.2}CuO_{4-y}F_y$ (circles) together with the $x$ dependence of $T_c$ in the hole-doped superconductors T'-$La_{1.8-x}Eu_{0.2}Sr_xCuO_4$ (squares)[10] and T'-$La_{1.8-x}Eu_{0.2}Ca_xCuO_4$ (diamonds).[19] The arrow indicates that the sample is not superconducting above 2 K. The dashed line indicates $T_c$ values calculated by Yamazaki et al.[24] on the basis of the spin-fluctuation-mediated pairing in the effective two-band model derived from the $d$-$p$ model, where the energy difference between the upper Hubbard level of the $Cu3d_{x^2-y^2}$ orbitals and the $O2p$ level, $\varDelta$, is put 1.43 eV and the $T_c$ values are normalized so that the maximum value of $T_c$ corresponds to the experimental one.